\documentclass[lettersize,journal]{IEEEtran}
\usepackage{graphicx}
\usepackage{graphics}
\usepackage{colortbl}
\usepackage{multicol}
\usepackage{diagbox}
\usepackage{lipsum}
\usepackage{color}
\usepackage[cmex10]{amsmath}
\usepackage{amsthm}
\usepackage{amssymb}
\usepackage{mathrsfs}
\usepackage{mathtools}
\usepackage{amsbsy}
\usepackage[colorlinks=true,bookmarks=false,citecolor=blue,urlcolor=blue]{hyperref}
\usepackage[dvipsnames]{xcolor}\usepackage{mathrsfs}
\usepackage{setspace}
\usepackage{float}
\usepackage{graphicx}
\usepackage{pstool}
\usepackage{epstopdf}
\usepackage{lettrine}
\usepackage{psfrag}
\usepackage{newclude}
\usepackage{dsfont}
\usepackage[normalem]{ulem}
\usepackage{latexsym}
\usepackage{algpseudocode}
\usepackage{algorithm,algpseudocode}
\usepackage{algorithmicx}
\usepackage{gensymb}
\usepackage{marginnote}
\usepackage{multirow}
\usepackage{amsmath}
\usepackage{amsmath,bm}
\usepackage{wasysym}
\usepackage{mathrsfs}
\usepackage{amsmath,cite,amsfonts,amssymb,color}
\usepackage[T1]{fontenc}
\usepackage{graphicx,psfrag}
\usepackage{pstool}
\usepackage{algorithm}
\usepackage{algpseudocode}
\usepackage{comment}
\usepackage{tabularx}
\usepackage{caption}
\usepackage{csquotes}
\usepackage{float}
\usepackage{lettrine}
\usepackage{makecell}
\usepackage{textcomp, gensymb}
\usepackage{relsize}
\usepackage{balance}
\usepackage{lipsum}

\ifCLASSOPTIONcompsoc
\usepackage[caption=false,font=normalsize,labelfon
t=sf,textfont=sf]{subfig}
\else
\usepackage[caption=false,font=footnotesize]{subfig}
\fi
\allowdisplaybreaks

\graphicspath{{figures/}}
\allowdisplaybreaks

\newcommand{\RNum}[1]{\uppercase\expandafter{\romannumeral #1\relax}}

\usepackage{mathtools}

\usepackage{comment}
\begin{document}

\title{\centering Robust Resource Allocation for LEO Satellite-Assisted Secure SWIPT via STAR-RIS under CSI Uncertainty}

\author{\IEEEauthorblockN{Zahra Rostamikafaki, Francois Chan, and  Claude D'Amours  \thanks{Zahra Rostamikafaki, Francois Chan, and  Claude D'Amours are with the Department of Electrical Engineering and Computer Science, University of Ottawa, Ottawa, ON K1N 6N5, Canada (E-mails: \{zrost034, cdamours,  fchan2\}@uottawa.ca).
Francois Chan is also with the Department of Electrical and Computer Engineering, Royal Military College of Canada, Kingston, ON K7K 7B4, Canada (e-mail: chan-f@rmc.ca).}
}}

% The paper headers
%\markboth{IEEE COMMUNICATIONS LETTERS,~Vol.~XX, No.~XX, 2023}%
%{Shell \MakeLowercase{\textit{et al.}}: A Sample Article Using IEEEtran.cls for IEEE Journals}

%\IEEEpubid{0000--0000/00\$00.00~\copyright~2021 IEEE}
% Remember, if you use this you must call \IEEEpubidadjcol in the second
% column for its text to clear the IEEEpubid mark.

\maketitle

\begin{abstract}
This paper proposes a robust resource allocation framework for a low Earth orbit (LEO) satellite-enabled simultaneous wireless information and power transfer (SWIPT) system, assisted by a ground-deployed simultaneously transmitting and reflecting reconfigurable intelligent surface (STAR-RIS). We consider a scenario where direct satellite-to-ground links are obstructed, and the satellite serves multiple single-antenna energy receivers, information receivers, and eavesdroppers exclusively via the STAR-RIS. A robust optimization problem is formulated to maximize the total harvested power, subject to secrecy rate requirements, transmit power limits, and STAR-RIS coefficient constraints, under a practical bounded channel state information (CSI) error model. To achieve optimal robust resource allocation,  we address the challenges posed by coupled optimization variables and bounded channel estimation errors by first applying the S-procedure to handle robustness against channel uncertainty. An alternating optimization (AO) framework is subsequently proposed, where the active beamforming at the LEO satellite and the passive beamforming at the STAR-RIS are jointly optimized, and a penalty-based strategy is incorporated to enforce the STAR-RIS beamforming design. Simulation results validate the effectiveness of the proposed algorithm and demonstrate that the STAR-RIS architecture achieves substantial performance gains in total harvested power over conventional RIS and other baseline schemes.
\end{abstract}


\section{Introduction}
\IEEEPARstart{T}{he} exponential growth of wireless connectivity, driven by advances in 5G and beyond, has led to the rapid proliferation of Internet of Things (IoT) devices across diverse sectors. Recent forecasts project that around 500 billion IoT devices will be connected worldwide within the next decade  \cite{cisco2018}, creating a massive ecosystem that enhances efficiency across numerous sectors but also introduces profound challenges in connectivity and energy provisioning. While terrestrial networks form the backbone of this expansion, they cannot economically serve remote and underserved regions, leaving approximately 4 billion people without reliable Internet access \cite{kota2019satellite}. To bridge this digital divide and support ubiquitous IoT, Low Earth Orbit (LEO) satellite constellations have emerged as a  key enabling technology, offering global coverage with significantly lower latency than their geostationary counterparts \cite{8002583},\cite{8417723}. LEO constellations provide disaster-resilient, secure backhaul and a uniform global service that keeps high-mobility assets connected across sparse terrestrial regions. The standardization of Non-Terrestrial Networks (NTN) by the Third Generation Partnership Project (3GPP) facilitates direct-to-device satellite communication using commercial off-the-shelf chipsets, thereby lowering costs and ensuring global interoperability \cite{3gppTR38811}. Specifically, the 5G New Radio (NR)-NTN protocol targets smartphones, while the Narrowband Internet of Things (NB-IoT)-NTN protocol is designed for massive IoT applications. Although these standards significantly advance direct satellite connectivity, major technical hurdles persist for handset-class links, including severe path loss, large Doppler shifts, and stringent timing-alignment constraints 
\cite{3gppTR36763}. To enhance LEO network performance, \cite{s22239304} proposes a multi-dimensional resource-allocation framework based on weighted greedy and genetic algorithms, while \cite{8951285} addresses fairness and reliability via an optimization approach using dual decomposition.

In parallel, ensuring long-term, self-sustainable operation of battery-constrained IoT nodes poses another major hurdle. Conventional energy sources such as batteries and wired connections are not only impractical for large-scale deployment but also result in high maintenance overhead and environmental concerns. To address this, wireless power transfer (WPT) and energy harvesting (EH) have gained increasing attention \cite{6489506}. In this context, simultaneous wireless information and power transfer (SWIPT) offers a compelling solution by enabling the joint transmission of information and energy over the same RF waveform \cite{6957150}. This facilitates concurrent service to information receivers (IRs) and energy harvesting receivers (EHRs)\cite{6951347}. However, a critical challenge lies in the disparity of power demands between EHRs and IRs where EHRs typically require significantly more RF energy to maintain functionality. The high free-space path loss and large distances in satellite-to-ground links pose serious challenges in delivering sufficient RF energy to EHRs. Initial studies have explored SWIPT in terrestrial networks \cite{6623062}, but applying it to the satellite domain amplifies these challenges due to the limited power budgets and extreme propagation distances involved. In satellite-enabled IoT environments, this challenge becomes even more acute due to the low-gain device antennas and intermittent visibility. Therefore, designing efficient SWIPT architectures that can maximize the harvested power at EHRs while satisfying information rate requirements and minimizing system complexity is essential for enabling scalable, energy-aware IoT connectivity via LEO satellite systems.

To address the limitations of traditional wireless propagation environments, reconfigurable intelligent surfaces (RIS) have recently emerged as a promising solution for realizing smart radio environments (SREs) in a cost-effective and energy-efficient manner \cite{9140329}. A RIS consists of a large array of low-cost, passive elements, each capable of independently modifying the phase of incoming electromagnetic waves. By controlling these phase shifts, RIS can create additional reflection paths to improve both signal coverage and quality. \cite{10313112},\cite{ 9134962 }. Integrating RIS into SWIPT systems has been shown to improve information throughput and harvested energy simultaneously \cite{9531372}, \cite{9623452}. Active RIS use amplification to reduce signal loss, which is especially helpful for long-distance or high-loss situations \cite{10155421 }. However, a major structural flaw of all conventional RIS both passive and active is that they are unidirectional, meaning transmitting and receiving devices must be placed on the same reflective side. This inherently limits deployment flexibility, especially in complex network topologies involving spatially distributed IoT devices.

To remedy this limitation, the simultaneously transmitting and reflecting reconfigurable intelligent surface (STAR-RIS) has been introduced \cite{xu2021star}. Unlike traditional RIS, which only reflects incident signals, STAR-RIS is composed of transparent elements that can independently manage both transmission and reflection, enabling 360-degree signal coverage \cite{liu2021star}. This expanded functionality allows STAR-RIS to serve users located on both sides of the surface, significantly enhancing deployment flexibility and improving system performance \cite{9570143}. Given these advantages, there is growing interest in integrating SWIPT systems with STAR-RIS, aiming to jointly enhance energy harvesting and information transfer across broader spatial domains \cite{10086660,10073379}.

Due to the inherent variability of wireless channels and imperfections in channel estimation techniques, acquiring perfect instantaneous channel state information (CSI) is generally infeasible \cite{7090935}. This issue is more pronounced in RIS-assisted systems, where the lack of active RF components prevents direct pilot transmission, making accurate estimation of cascaded channels particularly challenging. To tackle these limitations, robust design frameworks have been developed, such as the RIS-aided MISO system in \cite{9180053}, which considers both bounded and statistical CSI error models. These challenges also extend to STAR-RIS architectures, where simultaneous transmission and reflection operations introduce additional complexity in channel modeling and estimation. Recent works have begun to incorporate robust optimization techniques in STAR-RIS-aided SWIPT systems under CSI uncertainty, such as the bounded CSI model addressed in \cite{10859274,10525073}, highlighting the growing need for reliable CSI-aware design in next-generation wireless networks.

%To the best of our knowledge, the joint design of communication and power transfer for a STAR-RIS-aided LEO satellite system has not yet been investigated. This paper aims to fill this critical research gap. Specifically, we propose a novel framework that leverages a STAR-RIS to assist a LEO satellite in serving both IRs and EHRs distributed in its transmission and reflection spaces.

\subsection{Prior works}
\textit{1) RIS and STAR-RIS-Assisted SWIPT Systems:} 
Extensive research has demonstrated that integrating RIS and more recently, STAR-RIS into SWIPT systems can significantly improve both information transmission and energy harvesting performance \cite{8941080,9133120,9810984, 10244159,10347404,10918637, 10525073,10702481,10304608 }. 
In \cite{8941080}, the authors aimed to enhance the total harvested energy by jointly optimizing the AP's transmit precoding and the RIS phase configuration. Building on this, \cite{9133120} addressed energy efficiency maximization under secrecy constraints through coordinated design of beamforming and RIS control. To address the severe attenuation caused by double path loss, \cite{9810984} investigated the integration of active RIS with amplification capability. Numerous studies have demonstrated the capability of STAR-RIS to enhance spectral efficiency, energy efficiency, and communication security in wireless networks \cite{10244159,10347404,10918637}. Author in \cite{10525073} develops a robust transmit power minimization framework for a STAR-RIS-aided secure SWIPT system, jointly optimizing AP beamforming and STAR-RIS coefficients under secrecy and EH constraints with bounded CSI errors. A weighted sum power maximization problem was formulated for STAR-RIS-aided SWIPT systems in \cite{10702481}, where three practical STAR-RIS protocols energy splitting (ES), time switching (TS), and mode switching (MS) were employed to enhance the total harvested power at energy harvesting users (EHUs). In \cite{10304608}, the authors examined the balance between harvested power and information transmission rate in STAR-RIS-assisted SWIPT systems by employing ES and TS protocols to improve overall system performance.

\textit{2) LEO Satellite-Enabled SWIPT Systems:} 
A limited but growing number of studies from both academia and industry have investigated the integration of LEO satellites into SWIPT-enabled communication networks. For instance, the authors in \cite{zarini2024long} aim to jointly optimize the satellite’s downlink transmit power and the power splitting ratio at the devices, while ensuring queue stability, sufficient energy harvesting, and data rate requirements. In \cite{8720772}, the authors explore the use of SWIPT in satellite-ground integrated networks for LEO-based emergency communications, highlighting how power splitting effectively balances energy harvesting and data transmission to enhance overall communication efficiency. 

6G LEO satellite networks face significant challenges due to weak line-of-sight (LOS) conditions in dense urban environments, limited transmit power, and interference. RIS, known for their power efficiency and passive operation, offer a promising solution by enabling reliable reflected links that bypass obstacles and improve overall network coverage \cite{10559954}. In \cite{10008708}, the authors proposed a novel framework to improve the average throughput of RIS-assisted LEO networks by jointly optimizing RIS orientation and passive beamforming. The
authors of \cite{9621010} optimized both active and passive beamforming to enhance the received signal-to-noise ratio in RIS-assisted LEO satellite networks. Authors in \cite{khan2023ris} propose an energy-efficient RIS-assisted downlink NOMA communication framework for LEO satellite networks, addressing challenges such as high power consumption and spectral congestion. They jointly optimize the satellite's transmit power and RIS passive beamforming.

Owing to STAR-RIS's ability to provide full-space coverage via simultaneous reflection and transmission, it offers more effective support for LEO satellite communications than conventional RIS, by overcoming blockage and extending signal reach to users on both sides. For instance, authors in \cite{10702554} proposed an energy-efficient framework where STAR-RIS is mounted on UAVs to assist LEO satellite-based IoT communications. They jointly optimized the UAV trajectory, STAR-RIS phase shifts, and power allocation to maximize energy efficiency while ensuring fairness among users. A comprehensive study analyzed a STAR-RIS-assisted satellite–autonomous aerial vehicle–terrestrial NOMA system under the realistic condition of imperfect CSI, where outage probabilities were derived for multiple random deployment scenarios to capture the mobility effects in LEO satellite communications \cite{10891159}.  % In \cite{9726800}, the authors study a hybrid satellite-terrestrial relay network assisted by a refracting RIS and a half-duplex decode-and-forward base station, aiming to improve signal delivery to users with blocked direct links. They develop a joint optimization framework for beamforming and RIS phase shifts to minimize the overall transmit power while meeting user rate requirements.

While a few studies have investigated STAR-RIS-aided terrestrial SWIPT systems, to the best of our knowledge, no prior work has examined the integration of STAR-RIS with LEO satellites for SWIPT applications. This presents a significant and timely research gap, especially in light of the growing demand for energy-efficient and secure communications in future satellite-enabled networks. The fusion of LEO satellites with STAR-RIS enables full-space coverage and overcomes LOS blockages, making it ideal for dense urban areas. However, the omnidirectional transmission increases the risk of eavesdropping, necessitating robust security-aware designs. Unlike previous efforts that address either secrecy or energy performance in isolation, we provide a robust solution that accounts for both, specifically tailored for LEO satellite networks with STAR-RIS support. 

Our primary emphasis lies in jointly addressing the secrecy and energy harvesting challenges in SWIPT systems under bounded CSI uncertainty. Moreover, despite the advancements in RIS/STAR-RIS-assisted SWIPT systems under the assumption of perfect CSI, most existing studies remain confined to terrestrial networks \cite{8941080},\cite{10702481 }. Even in scenarios considering imperfect CSI, many works simplify the problem by either assuming common channels for eavesdroppers and energy harvesters or treating the eavesdropper as an energy-harvesting node \cite{10525073}. In contrast, our work tackles a more practical and challenging scenario by incorporating bounded CSI uncertainty and assigning distinct channels to each information user, energy harvester, and eavesdropper. Furthermore, while prior efforts such as \cite{khan2023ris} examined RIS-aided systems with a single-antenna LEO satellite for energy efficiency, the fundamental question of how effectively LEO satellites can support multiple ground energy harvesters in realistic SWIPT networks remains open. To address this, we consider a more practical setup with a multi-antenna LEO satellite, enabling flexible beamforming and joint optimization of active beamforming at the satellite and passive beamforming at the STAR-RIS, thereby offering new insights into the performance of LEO satellite-enabled SWIPT systems. 

%As highlighted earlier, although several important studies have advanced the understanding of RIS/STAR-RIS-assisted wireless networks under imperfect CSI, our work introduces a distinct perspective by focusing on the integration of STAR-RIS in LEO satellite-enabled SWIPT systems. Unlike prior research on terrestrial RIS-aided SWIPT systems with imperfect CSI, our emphasis lies in addressing the dual challenges of secure communication and efficient energy harvesting in LEO satellite networks. While [ ] investigated energy efficiency in RIS-aided systems with a single-antenna LEO satellite, our work considers a more generalized and practical multiple-input single-output (MISO) LEO scenario, enabling flexible beamforming capabilities. Furthermore, whereas earlier efforts often relied on statistical CSI or assumed ideal channel conditions, we incorporate bounded CSI uncertainty to design a robust optimization framework that guarantees both secrecy and energy efficiency in realistic LEO satellite settings.

Building upon the aforementioned motivations, this paper investigates a STAR-RIS-aided multiple-input single-output (MISO) system in LEO satellite-enabled SWIPT networks. Accordingly, our key contributions are outlined as follows:

\begin{itemize}
    \item We investigate a novel STAR-RIS-assisted SWIPT system in LEO satellite networks based on the channel bounded errors, where the satellite communicates exclusively via a ground-deployed STAR-RIS to serve IRs, EHRs, and potential eavesdroppers under blocked direct links.

    \item A joint robust beamforming framework is proposed to maximize the total harvested power at EHRs, which is critical for enabling sustainable operation in future 6G and IoT networks, while simultaneously ensuring secrecy rate requirements and power budget of the LEO satellite.

    \item To address the non-convex harvested-power maximization problem, we develop an efficient alternating optimization (AO) algorithm. The S-procedure is employed to transform the semi-infinite harvested power and secrecy constraints under bounded CSI uncertainty into tractable linear matrix inequalities (LMIs). The reformulated problem is then decomposed into two subproblems: active beamforming design at the multi-antenna LEO satellite and passive beamforming design at the ground-deployed STAR-RIS, which are solved in an alternating manner. Furthermore, a penalty-based method is incorporated to handle the rank-one constraints in the STAR-RIS optimization.

    \item Simulation results validate the effectiveness of the proposed framework and provide valuable insights into the STAR-RIS-aided SWIPT system. In particular, STAR-RIS significantly improves the sum harvested power at the energy receivers compared with conventional RIS and STAR-RIS schemes using equal energy split ratios, while also revealing the trade-offs between secrecy provisioning and power transfer efficiency in LEO satellite–integrated architectures. Moreover, evaluating discrete phase configurations under varying numbers of STAR-RIS elements and analyzing the system performance without secrecy constraints offers practical insights into real-world implementation.
\end{itemize}

The rest of the paper is organized as follows: Section II introduces the system model and formulates the robust harvested-power maximization problem for a STAR-RIS-aided LEO satellite SWIPT network. Section III presents the efficient solution for resource allocation under bounded CSI uncertainty. Section IV reports numerical results that demonstrate the performance gains of STAR-RIS compared with conventional baselines. Finally, Section V concludes this study.

\textit{Notation:} Matrices are denoted by bold uppercase letters, and vectors by bold lowercase letters. The transpose and conjugate transpose of a matrix $\mathbf{A}$ are represented by $\mathbf{A}^T$ and $\mathbf{A}^H$, respectively. The expectation operator is denoted by $\mathbb{E}\{\cdot\}$. The function $\mathrm{diag}(\mathbf{a})$ produces a diagonal matrix with the elements of vector $\mathbf{a}$ on its main diagonal. For a matrix $\mathbf{A}$, $\mathrm{Tr}(\mathbf{A})$ denotes its trace, $\mathrm{Rank}(\mathbf{A})$ its rank, and $\mathbf{A} \succeq 0$ indicates that $\mathbf{A}$ is positive semidefinite. The nuclear norm, spectral norm, and Frobenius norm of $\mathbf{A}$ are denoted by $\|\mathbf{A}\|_*$, $\|\mathbf{A}\|_2$, and $\|\mathbf{A}\|_F$, respectively. The Euclidean norm of a vector $\mathbf{a}$ is written as $\|\mathbf{a}\|$. The Kronecker product is represented by the symbol $\otimes$. Finally, the circularly symmetric complex Gaussian distribution with mean $\mu$ and variance $\sigma^2$ is denoted by $\mathcal{CN}(\mu, \sigma^2)$.

\begin{figure}[t!]
\centering
\includegraphics[width=0.45\textwidth]{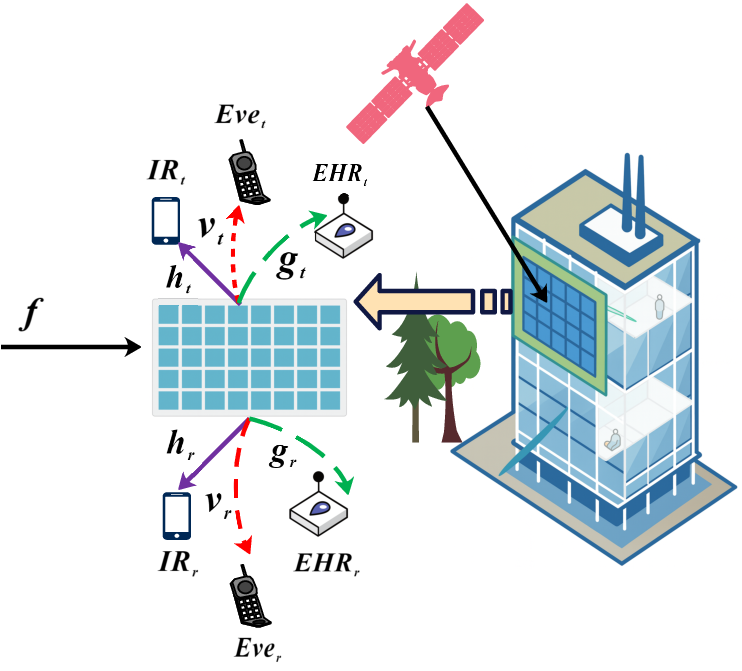}
    \caption{The STAR-RIS aided secure SWIPT system.}
    \label{fig:1}
\end{figure}

\section{System Model}
Fig.\ref{fig:1} depicts the downlink of a secure LEO–satellite MISO SWIPT system assisted by a STAR–RIS. A LEO satellite with an $N_t$-element transmit array serves three types of single-antenna ground terminals: information receivers (IRs), energy-harvesting receivers (EHRs), and eavesdroppers (Eves). Due to unfavorable propagation conditions such as terrain and environmental blockages the direct satellite-to-terminal links are unavailable; hence, communication and energy transfer are achieved solely via an $M$-element STAR-RIS that both reflects and transmits the incident signal toward the intended users. Terminals $\text{IR}_t$, $\text{EHR}_t$, and $\text{Eve}_t$ are located in the transmission region of the STAR--RIS, while $\text{IR}_r$, $\text{EHR}_r$, and $\text{Eve}_r$ reside in its reflection region. The IR in the $k$-th region is denoted $\text{IR}_k$, $\forall k\in\{t,r\}$; the same goes for the energy harvester and the eavesdropper, denoted $\text{EHR}_k$ and $\text{Eve}_k$, respectively.

The satellite to STAR-RIS channel is modeled as a line-of-sight (LoS), which includes satellite antenna pattern effects, Doppler shifts, and free-space path loss. The baseband equivalent LEO satellite-to-ground channel to STAR-RIS is represented as $f_k = \hat{f}_k \, e^{j\pi \upsilon_k}$ where \( \hat{f}_k \) denotes the complex-valued channel coefficient accounting for both large-scale and small-scale fading, and \( \upsilon_k \) is the Doppler shift, and \( j = \sqrt{-1} \). The magnitude \( \hat{f}_k \) is modeled as \cite{khan2023ris}:
\begin{equation}
\hat{f}_k = \frac{\sqrt{G_\ell \, G_{\ell,s}} \, c}{4\pi f_c d_l}
\label{eq:1}
\end{equation}
Here, \( c \) is the speed of light, \( f_c \) is the carrier frequency, and \( d_l \) represents the slant range distance between the LEO satellite and the STAR-RIS. The parameters \( G_\ell \) and \( G_{\ell,s} \) correspond to the satellite's transmit antenna gain and the effective receive gain at the STAR-RIS-facing side, respectively. The satellite’s antenna gain is dependent on its beam radiation pattern and is approximated by:
\begin{equation}
G_\ell = G_{\text{max}} \left( \frac{J_1(\vartheta_k)}{2\vartheta_k} + 36 \cdot \frac{J_3(\vartheta_k)}{\vartheta_k^3} \right)^2
\label{eq:2}
\end{equation}
where \( G_{\text{max}} \) is the maximum gain at the beam center, \( J_1(\cdot) \) and \( J_3(\cdot) \) are the first-kind Bessel functions of orders 1 and 3, respectively, and the normalized angular offset is defined as $\vartheta_k = \frac{2.07123 \sin(\theta_k)}{\sin(\theta_{\text{3dB}})}$
where \( \theta_k \) denotes the angular deviation of STAR-RIS and the center of LEO satellite beam, and \( \theta_{\text{3dB}} \) is the satellite antenna's 3 dB beamwidth. Doppler shift due to LEO motion is assumed to be fully compensated at the receiver.

We consider a STAR-RIS with $M$ elements, indexed by $m \in \mathcal{M} \triangleq \{1,2,\dots,M\}$. Each element simultaneously transmits and reflects the incident signal $x_m$ from the satellite, modeled as $t_m = \sqrt{\beta_m^t} e^{j\theta_m^t} x_m$ and $r_m = \sqrt{\beta_m^r} e^{j\theta_m^r} x_m$. Here, $\sqrt{\beta_m^t}, \sqrt{\beta_m^r} \in [0,1]$ denote the transmission and reflection amplitude coefficients, and $\theta_m^t, \theta_m^r \in [0, 2\pi)$ represent their respective phase shifts. 
The energy splitting (ES) protocol allows all STAR-RIS elements to operate in both modes simultaneously, offering design flexibility. The corresponding transmission and reflection coefficient matrices are defined as $\boldsymbol{\Theta}_t = \text{diag}(\sqrt{\beta_1^t}e^{j\theta_1^t}, \dots, \sqrt{\beta_M^t}e^{j\theta_M^t})$ and $\boldsymbol{\Theta}_r = \text{diag}(\sqrt{\beta_1^r}e^{j\theta_1^r}, \dots, \sqrt{\beta_M^r}e^{j\theta_M^r})$, where energy conservation per element is enforced by $\beta_m^t + \beta_m^r = 1$.

\subsection{Signal Transmission Model}
The satellite transmits a precoded signal $\mathbf{x}=\sum_{k} \mathbf{w}_k s_k$ , $k\in \{t,r\}$ to the each \text{IR} at the same frequency, where $\mathbf{w}_k\in\mathbb{C}^{N_t\times 1}$ is the transmit beamforming vector and $s_k$ is the information symbol intended for $\text{IR}_k$, satisfying $\mathbb{E}[|s_k|^2]=1$.

The received signal at the $\text{IR}_k$ is given by:
\begin{equation}
y_{I,k} = \mathbf{h}_{k}^H \boldsymbol{\Theta}_{k} \mathbf{f}  \sum_{k} \mathbf{w}_k s_k
 + n_{I,k},
\end{equation}
where $\mathbf{h}_{k}^H \in \mathbb{C}^{1 \times M}$, $\mathbf{f} \in \mathbb{C}^{M \times N_t}$, and $\boldsymbol{\theta}_{k} \in \mathbb{C}^{M \times M}$ represent the channels between STAR-RIS and the $\text{IR}_k$, the channel from the LEO satellite to the STAR-RIS, and STAR-RIS coefficient for $\text{IR}_k$, respectively. $n_{I,k} \sim \mathcal{CN}(0,\sigma_k^2)$ is the additive white Gaussian noise (AWGN).

The signal received by the $\text{Eve}_k$ is:
\begin{equation}
y_{e,k} =  \mathbf{v}_{k}^H \boldsymbol{\Theta}_{k} \mathbf{f} \sum_{k } \mathbf{w}_k s_k
 + n_{e,k},
\end{equation}

where $\mathbf{v}_k \in \mathbb{C}^{M \times 1}$ denotes the channel from the STAR-RIS to the $\text{Eve}_k$, and $n_{e,k} \sim \mathcal{CN}(0, \sigma_e^2)$ represents the AWGN at the eavesdropper.

The total harvested power at the $\text{EHR}_k$ is expressed as:
\begin{equation}
E_{k} = \eta_k \, \mathbb{E}\!\left[
\left| \mathbf{g}_{k}^H \boldsymbol{\Theta}_{k} \mathbf{f}
\sum_{k } \mathbf{w}_k s_k \right|^2
\right]
= \eta_k\sum_{k }
\left| \mathbf{g}_{k}^H \boldsymbol{\Theta}_{k} \mathbf{f} \mathbf{w}_k \right|^2,
\end{equation}

where $\eta_k\in (0, 1]$ is the energy conversion efficiency and $\mathbf{g}_{k} \in \mathbb{C}^{M \times 1}$ represent the channel between STAR-RIS and $\text{EHR}_k$.

Accordingly, the achievable rate at the $\text{IR}_k$ and the corresponding leakage rate at the $\text{Eve}_k$ in region $k \in \{t, r\}$ are:
\begin{equation}
R_{I,k} = \log_2 \left(1 + \frac{|\boldsymbol{\Phi}_{k}^H \mathbf{H}_{k} \mathbf{w}_k|^2}
{|\boldsymbol{\Phi}_{k}^H \mathbf{H}_{k} \mathbf{w}_{\acute{k}}|^2 + \sigma_k^2} \right),
\end{equation}
 
\begin{equation}
R_{e,k} = \log_2 \left(1 + \frac{|\boldsymbol{\Phi}_{l}^H \mathbf{V}_{k} \mathbf{w}_k|^2}
{ |\boldsymbol{\Phi}_{l}^H\mathbf{V}_{k} \mathbf{w}_{\acute{k}}|^2 + \sigma_e^2} \right).
\end{equation}
where \(\mathbf{\Phi}_k = \big[ \sqrt{\beta^{k}_1}e^{j\theta^{k}_1},\, \ldots,\, \sqrt{\beta^{k}_M}e^{j\theta^{k}_M} \big]^{H}\),
$\mathbf{H}_{k} = \text{diag}(\mathbf{h}_{k}^H)\mathbf{f}$ and $\mathbf{V}_{k} = \text{diag}(\mathbf{v}_{k}^H)\mathbf{f}$ respectively.
Moreover, if $k=t$ then $\acute{k}=r$, and the reverse holds. This model forms the foundation for secure and efficient resource allocation in STAR-RIS-assisted LEO satellite SWIPT systems.

\subsection{CSI Error Model}
Due to the passive characteristics and complex propagation environment associated with STAR-RIS-assisted systems, acquiring perfect channel state information (CSI) is highly impractical for the links between the STAR-RIS and the ground users. In this work, we assume that the satellite-to-STAR-RIS channel is line-of-sight (LoS) and perfectly known, while the STAR-RIS-to-user channels are subject to estimation errors. To capture the impact of imperfect CSI in the beamforming design, we adopt a bounded CSI error model for the cascaded channels from the LEO satellite to each user through the STAR-RIS.

Specifically, the cascaded channels are modeled as:
\begin{align}
\mathbf{H}_k &= \hat{\mathbf{H}}_k + \Delta \mathbf{H}_k, \quad \forall k \in \{t, r\}, \\
\mathcal{H}_k &= \left\{ \Delta \mathbf{H}_k \in \mathbb{C}^{M \times N_t} : \| \Delta \mathbf{H}_k \|_F \leq \epsilon_{1,k} \right\}, \\
\mathbf{G}_k &= \hat{\mathbf{G}}_k + \Delta \mathbf{G}_k, \quad \forall k \in \{t, r\}, \\
\mathcal{G}_k &= \left\{ \Delta \mathbf{G}_k \in \mathbb{C}^{M \times N_t} : \| \Delta \mathbf{G}_k \|_F \leq \epsilon_{2,k} \right\}, \\
\mathbf{V}_k &= \hat{\mathbf{V}}_k + \Delta \mathbf{V}_k, \quad \forall k \in \{t, r\}, \\
\mathcal{V}_k &= \left\{ \Delta \mathbf{V}_k \in \mathbb{C}^{M \times N_t} : \| \Delta \mathbf{V}_k \|_F \leq \epsilon_{3,k}  \right\}.
\end{align}

Here, \( \hat{\mathbf{H}}_k \), \( \hat{\mathbf{G}}_k \), and \( \hat{\mathbf{V}}_k \) represent the estimated effective cascaded channels from the satellite to the \(k\)-th information receiver, energy harvesting receiver, and eavesdropper, respectively. The matrices \( \Delta \mathbf{H}_k \), \( \Delta \mathbf{G}_k \), and \( \Delta \mathbf{V}_k \) denote the corresponding channel estimation errors, each bounded in Frobenius norm by predefined constants \( \epsilon_{1,k} \), \( \epsilon_{2,k} \), and \( \epsilon_{3,k} \), respectively. This uncertainty model enables robust beamforming optimization under worst-case CSI deviations.

\subsection{Problem Formulation}
In this work, we aim to maximize the total harvested power across all $\text{EHR}_k$ while simultaneously ensuring that the secrecy rate constraints of $\text{IR}_k$ are met in the presence of eavesdroppers. To achieve this, we jointly optimize the active beamforming vectors \( \mathbf{w}_k \) and the passive beamforming matrix \( \boldsymbol{\Phi}_l \). Considering the presence of CSI uncertainty, we formulate a robust optimization problem accordingly.

\begin{subequations} \label{eq:14}
\begin{align}
\max_{\mathbf{w}_k, \boldsymbol{\Phi}_k} \quad & \sum_{k} \text{E}_k , \quad \forall k \in \{t, r\} \label{eq:14a}\\
\text{s.t.} \quad 
& R_{I,k} \geq R_{\text{th}}^{\text{IR}}, \quad \forall k  \label{eq:14b}\\
& R_{e,k} \leq R_{\text{th}}^{\text{Eve}}, \quad \forall k \label{eq:14c} \\
& \sum_{k} {\big| \boldsymbol{w}_k \big|}^2 \le P_{\text{max}}, \quad \forall k \label{eq:14d} \\
& \Phi_k \in [0, 2\pi], \quad \forall k \in \{t, r\} \label{eq:14e}\\
& \beta^t_m + \beta^r_m = 1, 0 \leq \beta^{t/r}_m \leq 1
\quad \forall m \in \mathcal{M}\label{eq:14f}
\end{align}
\end{subequations}

Where \( R_{\text{th}}^{IR} \) is the minimum required rate threshold for each IR, \( R_{th}^{\text{Eve}} \) is the maximum tolerable leakage rate at the eavesdroppers. Constraint \eqref{eq:14a} defines the objective to maximize the total harvested power. Constraint \eqref{eq:14d} represents the transmit power limit, Constraint \eqref{eq:14e} constrains the STAR-RIS phase shifts, and Constraint \eqref{eq:14f} enforces energy conservation between transmission and reflection coefficients.

Note that problem \eqref{eq:14} is a non-convex and intractable optimization problem, which cannot be solved directly in its current form. This complexity arises for two main reasons. First, the optimization variables, namely the active beamforming vectors \( \mathbf{w}_k \) and the STAR-RIS coefficients \( \boldsymbol{\Phi}_l \), are intricately coupled. Second, the presence of imperfect CSI introduces uncertainty into both the objective function and constraints (\ref{eq:14b}) and (\ref{eq:14c}), leading to infinitely many constraint realizations. To address these challenges, we develop an efficient iterative algorithm based on the alternating optimization (AO) framework to obtain a high-quality suboptimal solution.

\section{Proposed Solution}
To tackle the non-convex optimization problem in \eqref{eq:14}, we adopt a two-stage alternating optimization (AO) framework. First, the original problem is equivalently reformulated into a rank-constrained semidefinite program (SDP). Leveraging the bounded channel uncertainty, the S-procedure is applied to convert the robust harvested power and secrecy rate constraints into linear matrix inequalities (LMIs), enabling tractable optimization over the imperfect CSI domain.

\subsection{Problem Reformulation}
To facilitate a tractable reformulation, we define the transmit beamforming matrix as \(\mathbf{W}_k = \mathbf{w}_k \mathbf{w}_k^H\), \(\forall k \in \{t, r\}\), which satisfies the positive semidefinite constraint \(\mathbf{W}_k \succeq 0\) and \(\text{Rank}(\mathbf{W}_k) = 1\). Moreover, the STAR-RIS phase shift matrix on side $k \in \{t, r\}$ is modeled as $\mathbf{Q}_k = \boldsymbol{\Phi}_k \boldsymbol{\Phi}_k^H$, where $\boldsymbol{\Phi}_k \in \mathbb{C}^{M \times 1}$ is a complex vector representing the reflection/transmission coefficients. Hence, $\mathbf{Q}_k$ is Hermitian positive semidefinite and rank-one with $\text{diag}(\mathbf{Q}_k) = \boldsymbol{\beta}_k$, where $\boldsymbol{\beta}_k = [\beta_{k,1}, \beta_{k,2}, \dots, \beta_{k,M}]$. 

To facilitate tractable reformulation under imperfect CSI, we introduce an auxiliary variable $\xi$ such that $\sum_{k} \text{E}_k \geq \xi, \ \forall k \in  \{t,r\}$. This allows the robust optimization problem to be equivalently reformulated.

\begin{subequations} \label{eq:15}
\begin{align} 
\quad  \max_{\{\mathbf{W}_k\}, \mathbf{Q}_k, \xi} \quad & \xi \notag \\
\text{s.t.} \quad 
& \mathrm{Tr}\left( \left( \sum_{k } \mathbf{W}_k \right) 
\mathbf{G}_k^H \mathbf{Q}_k \mathbf{G}_k \right) \geq \xi, \nonumber \\
& \quad \forall k \in \{t, r\}, \label{15a} \\
& \mathrm{Tr} \left( \left( \frac{\mathbf{W}_k}{\Gamma} 
- \mathbf{W}_{\acute{k}} \right) 
\mathbf{H}_k^H \mathbf{Q}_{k} \mathbf{H}_k \right) \geq \sigma_k^2, \nonumber \\
& \quad \forall \ k \in \{t, r\}, \label{15b} \\
&\mathrm{Tr}\left( \left(  \mathbf{W}_{\acute{k}} - \frac{\mathbf{W}_k}{\eta} \right) 
\mathbf{V}_k^H \mathbf{Q}_k \mathbf{V}_k \right) + \sigma_e^2 \geq 0 \notag \\
&\quad \forall \ k \in \{t, r\}, \label{15c} \\
& \sum_{k } \mathrm{Tr}(\mathbf{W}_k) \leq P_{\max}, \label{15d} \\
& \mathbf{W}_k \succeq 0,\ \operatorname{Rank}(\mathbf{W}_k) = 1, 
\quad \forall k , \label{15e} \\
& \mathbf{Q}_k \succeq 0,\ \operatorname{Rank}(\mathbf{Q}_k) = 1, 
\quad \forall k , \label{15f} \\
& \mathrm{diag}(\mathbf{Q}_k) = \boldsymbol{\beta}_k, 
\quad \forall k , \label{15g} \\
& \beta^t_m + \beta^r_m = 1, 0 \leq \beta^{t/r}_m \leq 1
\quad \forall m \in \mathcal{M}.\label{15h}
\end{align}
\end{subequations}

where $\Gamma = 2^{R_{\text{th}}^{\text{IR}}} - 1$ and $\eta = 2^{R_{\text{th}}^{\text{Eve}}} - 1$ denote the SINR thresholds corresponding to the minimum rate requirements of the IRs and the maximum tolerable rate of the eavesdroppers, respectively.

Using the identity $\mathrm{Tr}(\mathbf{A}^H \mathbf{B} \mathbf{C} \mathbf{D}) = \operatorname{vec}(\mathbf{A})^H (\mathbf{D}^T \otimes \mathbf{B}) \operatorname{vec}(\mathbf{C})$, 
The constraints in \eqref{15a}, \eqref{15b}, and \eqref{15c} can be equivalently expressed in a vectorized form suitable for robust optimization.

\begin{equation}
\operatorname{vec}(\mathbf{G}_k)^{H}\!\left(\sum_{k }\mathbf{W}_k^{T}\otimes \mathbf{Q}_k\right)\!\operatorname{vec}(\mathbf{G}_k)\ge \xi,\quad \forall\, \; k\in\{t,r\},
\label{16}
\end{equation}
 
\begin{align}
\operatorname{vec}(\mathbf{H}_k)^H 
\left( \left( \frac{\mathbf{W}_k}{\Gamma} -  \mathbf{W}_{\acute{k}} \right)^T 
\otimes \mathbf{Q}_k \right) 
\operatorname{vec}(\mathbf{H}_k) &\geq \sigma_k^2, \notag \\
\forall \ k \in \{t, r\}, && \label{17} \\[0.5em]
\operatorname{vec}(\mathbf{V}_k)^H 
\left( \left(  \mathbf{W}_{\acute{k}} 
- \frac{\mathbf{W}_k}{\eta} \right)^T \otimes \mathbf{Q}_k \right) 
\operatorname{vec}(\mathbf{V}_k) + \sigma_e^2 &\geq 0, \notag \\
,\forall \ k \in \{t, r\}. && \label{18}
\end{align}

To address the channel uncertainty present in equations \eqref{16},\eqref{17}, and \eqref{18} we apply the following result:

\textit{Lemma 1 (Generalized S-Procedure \cite{boyd2004convex}):}  
\textit {Consider a set of quadratic functions defined over the complex vector \( \mathbf{x} \in \mathbb{C}^{N \times 1} \) as  
\[
f_k(\mathbf{x}) = \mathbf{x}^H \mathbf{A}_k \mathbf{x} + 2\Re\{\mathbf{x}^H \mathbf{b}_k\} + c_k, \quad \forall k = 1, \dots, K,
\]  
where \( \mathbf{A}_k \in \mathbb{C}^{N \times N} \) are Hermitian matrices (i.e., \( \mathbf{A}_k = \mathbf{A}_k^H \)).  
Then, the implication \( f_k(\mathbf{x}) \geq 0, \forall k = 1,\dots,K \Rightarrow f_0(\mathbf{x}) \geq 0 \) holds \textit{if and only if} there exist non-negative scalars \( \lambda_k \geq 0 \) such that the following matrix inequality is satisfied:}

\[
\begin{bmatrix}
\mathbf{A}_0 & \mathbf{b}_0 \\
\mathbf{b}_0^H & c_0
\end{bmatrix}
- \sum_{k=1}^{K} \lambda_k
\begin{bmatrix}
\mathbf{A}_k & \mathbf{b}_k \\
\mathbf{b}_k^H & c_i
\end{bmatrix}
\succeq 0.
\]

Let the channel in \eqref{17} be expressed as $\mathbf{H}_k = \hat{\mathbf{H}}_k + \Delta \mathbf{H}_k$ where $\mathbf{S}_1 = \left( \frac{\mathbf{W}_k}{\Gamma} - \mathbf{W}_{\acute{k}} \right)^{T}$. 
By substituting into the constraint, we have
\begin{equation}
\operatorname{vec}(\hat{\mathbf{H}}_k + \Delta \mathbf{H}_k)^H 
\left( \mathbf{S}_1 \otimes \mathbf{Q}_{k} \right)
\operatorname{vec}(\hat{\mathbf{H}}_k + \Delta \mathbf{H}_k) - \sigma_k^2 \geq 0,
\label{eq:19}
\end{equation}
which is equivalent to
\begin{equation}
\operatorname{vec}(\Delta \mathbf{H}_k)^H \mathbf{A}_1 \operatorname{vec}(\Delta \mathbf{H}_k) 
+ 2 \Re\left\{ \operatorname{vec}(\Delta \mathbf{H}_k)^H \mathbf{b} \right\} + c \geq 0,
\label{eq:20}
\end{equation}
Here, we define $\mathbf{A}_1 = \mathbf{S}_1 \otimes \mathbf{Q}_{k}$, 
$\mathbf{b} = \mathbf{A}_1 \operatorname{vec}(\hat{\mathbf{H}}_k)$, 
and $c = \operatorname{vec}(\hat{\mathbf{H}}_k)^H \mathbf{b} - \sigma_k^2$.

Furthermore, since $\|\Delta\mathbf{H}_k\|_F\leq\epsilon_{1,k}$ implies $\|\operatorname{vec}(\Delta \mathbf{H}_k)\|_2 \leq \epsilon_{1,k}$, the channel uncertainty constraint can be reformulated as the following quadratic constraint:
\begin{equation}
-\operatorname{vec}(\Delta \mathbf{H}_k)^H \mathbf{I} \operatorname{vec}(\Delta \mathbf{H}_k) + \epsilon_{1,k}^2 \geq 0.
\label{eq:21}
\end{equation}

\begin{equation} \label{eq:22}
\begin{pmatrix}
\mathbf{S}_1 \otimes \mathbf{Q}_k + \lambda_{1,k} \mathbf{I} & (\mathbf{S}_1 \otimes \mathbf{Q}_k)\, \operatorname{vec}(\hat{\mathbf{H}}_k) \\
\operatorname{vec}(\hat{\mathbf{H}}_k)^H\, (\mathbf{S}_1 \otimes \mathbf{Q}_k) & \alpha_k
\end{pmatrix}
\succeq 0
\end{equation}
where \( \alpha_k = \operatorname{vec}(\hat{\mathbf{H}}_k)^H (\mathbf{S}_1 \otimes \mathbf{Q}_k) \operatorname{vec}(\hat{\mathbf{H}}_k) - \sigma_k^2 - \lambda_{1,k}  \epsilon_{1,k}^2 \), \( \lambda_{1,k} \geq 0 \) is an auxiliary variable introduced by the S-procedure, and \( \epsilon_{1,k} \) denotes the uncertainty bound. This formulation transforms the original quadratic inequality with bounded CSI error into an LMI, enabling convex optimization-based robust beamforming design.

Likewise, constraints \eqref{16} and \eqref{18} are converting to LMI form as: 
 
\begin{equation} \label{eq:23}
\begin{pmatrix}
\mathbf{S}_{2,k}\otimes \mathbf{Q}_k + \lambda_{2,k} \mathbf{I} & (\mathbf{S}_{2,k}\otimes \mathbf{Q}_k) \operatorname{vec}(\hat{\mathbf{G}}_k) \\
\operatorname{vec}(\hat{\mathbf{G}}_k)^H (\mathbf{S}_{2,k}\otimes \mathbf{Q}_k) & \kappa_k
\end{pmatrix}
\succeq 0,
\end{equation}
where  $\mathbf{S}_{2,k} = \sum_{k }\mathbf{W}_k^T$, $\kappa_k = \operatorname{vec}(\hat{\mathbf{G}}_k)^H (\mathbf{S}_{2,k}\otimes \mathbf{Q}_k) \operatorname{vec}(\hat{\mathbf{G}}_k) - \xi - \lambda_{2,k} \epsilon_{2,k}^2$, and $\lambda_{2,k} \geq 0$ is an auxiliary variable.

\begin{equation} \label{eq:24}
\begin{pmatrix}
\mathbf{S}_{3,k} \otimes \mathbf{Q}_k + \lambda_{3,k} \mathbf{I} & (\mathbf{S}_{3,k} \otimes \mathbf{Q}_k) \operatorname{vec}(\hat{\mathbf{V}}_k) \\
\operatorname{vec}(\hat{\mathbf{V}}_k)^H (\mathbf{S}_{3,k} \otimes \mathbf{Q}_k) & \varrho_k
\end{pmatrix}
\succeq 0
\end{equation}
where \( \mathbf{S}_{3,k} = \left( \mathbf{W}_{\acute{k}} - \frac{\mathbf{W}_k}{\eta} \right)^T \), $\lambda_{3,k} \geq 0$, and 
\[
\varrho_k = \operatorname{vec}(\hat{\mathbf{V}}_k)^H (\mathbf{S}_{3,k}^T \otimes \mathbf{Q}_k) \operatorname{vec}(\hat{\mathbf{V}}_k) + \sigma_e^2 - \lambda_{3,k} \epsilon_{3,k}^2,
\]

after these transformation the problem \eqref{eq:14} is recasted as: 
\begin{subequations} \label{eq:25}
\begin{align}
\max_{\{\lambda_{1,k}\}, \{\lambda_{2,k}\}, \{\lambda_{3,k}\}, \{\xi\}, \{\mathbf{Q}_k\}, \{\mathbf{W}_k\}} \quad & \xi \notag \\
\text{s.t.} \quad
\eqref{eq:22}, \eqref{eq:23}, \eqref{eq:24},\eqref{15d}-\eqref{15h}\label{eq:25a}.\\ \notag
\end{align}
\end{subequations}

It is evident that the optimization variables in \eqref{eq:22}, \eqref{eq:23}, \eqref{eq:24} remain coupled, making their joint optimization a nontrivial task. To address this complexity, we begin by splitting problem \eqref{eq:25} into two separate components: the design of active beamforming and the design of passive beamforming. These components are then optimized in an alternating manner using the proposed AO-based approach.
\subsection{Active Beamforming Design}
In this part, we focus on the design of active beamforming. Assuming that $\{\mathbf{Q}_k\}$ are fixed, problem \eqref{eq:25} can be reformulated as:
\begin{subequations} \label{eq:26}
\begin{align}
\max_{\{\lambda_{1,k}\}, \{\lambda_{2,k}\}, \{\lambda_{3,k}\},\{\mathbf{W}_k\}, \{\xi\} } \quad & \xi \notag \\
\text{s.t.} \quad
\eqref{15d},\eqref{15e},\eqref{eq:22}, \eqref{eq:23}, \eqref{eq:24}\label{eq:26a}.\\ \notag
\end{align}
\end{subequations}

The core challenge in resolving problem \eqref{eq:26} is due to its non-convexity, which arises from the rank-one constraints in \eqref{15e}. To address this issue, we apply the semi-definite relaxation (SDR) method by dropping the rank-one constraints, resulting in a convex optimization problem. The effectiveness of this relaxation is established through the following theorem \cite{10304608}.

\textit{Theorem 1: The optimal solutions for the relaxed version of problem \eqref{eq:26}, i.e., without considering rank-one constraints, always satisfy $\operatorname{Rank}(W_k^\star) = 1,\ \forall\, k \in \{t, r\}$, for any feasible instance with $P_{\max}>0$ and $\Gamma > 0, \eta > 0$.}

\emph{Proof: Please refer to Appendix~1.}

The resulting relaxed formulation corresponds to a conventional SDP, which can be efficiently addressed using well-established convex optimization tools, such as CVX \cite{boyd2004convex}.

\subsection{Passive Beamforming Design}
In this stage, the focus is on optimizing the passive beamforming while keeping 
$\{W_k\}$ fixed. The problem can be reformulated as:
\begin{subequations} \label{eq:27}
\begin{align}
\max_{\{\lambda_{1,k}\}, \{\lambda_{2,k}\}, \{\lambda_{3,k}\},\{\mathbf{Q}_k\},\{\xi\}} \quad  \xi \notag \\
\text{s.t.} \quad
\eqref{15f}-\eqref{15h},\eqref{eq:22}, \eqref{eq:23}, \eqref{eq:24}\label{eq:27a}.\\ \notag
\end{align}
\end{subequations}

The main challenge in 
determining the STAR-RIS coefficients lies in satisfying the rank-one condition 
in \eqref{15f}. Unlike the approach taken in the relaxation of problem \eqref{eq:26}, we begin 
by expressing this condition in an equivalent form as:

\begin{equation}
\|\mathbf{Q}_k\|_* - \|\mathbf{Q}_k\|_2 = 0, \quad \forall k \in \{t, r\},
\label{eq:28}
\end{equation}
where $\|\mathbf{Q}_l\|_*$ and $\|\mathbf{Q}_k\|_2$ denote the nuclear norm and spectral norm of $\mathbf{Q}_k$, respectively. Since $\mathbf{Q}_k \succeq 0$, the inequality $\|\mathbf{Q}_k\|_* - \|\mathbf{Q}_k\|_2 \geq 0$ always holds, with equality if and only if $\text{Rank}(\mathbf{Q}_k) = 1$.

To enforce this rank condition, we introduce a penalty term with factor $\gamma > 0$ in the objective function. Using SCA, the upper bound of the penalty term is approximated as:
\begin{equation}
\|\mathbf{Q}_k\|_* - \|\mathbf{Q}_k\|_2 \leq \|\mathbf{Q}_k\|_* - \mathbf{Q}_k^{(i)},
\label{eq:29}
\end{equation}
where
\begin{equation}
 \mathbf{Q}_k^{(i)} = \|\mathbf{Q}_k^{(i)}\|_2 + \text{Tr} \left[ \mathbf{u} (\mathbf{Q}_k^{(i)}) (\mathbf{u} (\mathbf{Q}_k^{(i)}))^H (\mathbf{Q}_k - \mathbf{Q}_k^{(i)}) \right],
\end{equation}
and $\mathbf{u} (\mathbf{Q}_k^{(i)})$ is the dominant eigenvector of $\mathbf{Q}_k$.

Given the penalty term approximation, the problem is reformulated as:

\begin{subequations} \label{eq:31}
\begin{align}
\quad \max_{\{\lambda_{1,k}\}, \{\lambda_{2,k}\}, \{\lambda_{3,k}\}, \{\mathbf{Q}_k\},\{\xi\}} \quad & 
\xi - \gamma \sum_{l \in \{t, r\}} \left( \| \mathbf{Q}_k \|_* - \mathbf{Q}_k^{(i)} \right)\notag  \\
\text{s.t.} \quad 
&\eqref{15g},\ \eqref{15h},\ \eqref{eq:22},\ \eqref{eq:23},\ \eqref{eq:24}. \label{eq:31a} 
\end{align}
\end{subequations}
Here, it is worth noting that in this 
maximization problem, increasing the value of $\gamma$ parameter reduces the magnitude of the associated penalty term. In the limit, as $\gamma \to +\infty$, 
the obtained $\mathbf{Q}_k$ from problem \eqref{eq:31} inherently fulfills the equality 
constraint \eqref{15f}.

To solve this, a penalty-based iterative algorithm is adopted. In the outer loop, $\gamma$ is updated as \( \gamma^{(i+1)} = c \gamma^{(i)} \) with \( c > 1 \), while the inner loop solves the relaxed version of the problem. The algorithm terminates when:
\begin{equation}
\sum_{l \in \{r,t\}} (\|\mathbf{Q}_k\|_* - \mathbf{Q}_k^{(i)}) \leq \delta.
\label{eq:32}
\end{equation}

\begin{algorithm}[t]
\caption{Penalty-Based Algorithm for Solving Problem (31)}
\begin{algorithmic}[1]
\State Initial values for $\{\mathbf{Q}_k^{(0)}\}$, penalty factor $\gamma$, and given $\{\mathbf{W}_k\}$; set the tolerances $\delta_1$, $\delta_2$, and the maximum number of iterations $I_{\max}$.
\Repeat
    \State Set iteration index $I = 0$;
    \Repeat
        \State with the current $\{\mathbf{Q}_k^{(I)}\}$, solve the problem (31);
        \State Update $\{\mathbf{Q}_k^{(I+1)}\}$ with the obtained optimal solutions, $I = I + 1$;
    \Until{until the increase in the objective function becomes smaller than a predefined tolerance level $\delta_2 > 0$ or $I = I_{\max}$}
    \State Update $\{\mathbf{Q}_\ell^{(0)}\}$ with the optimized solutions $\{\mathbf{Q}_k^{(I)}\}$;
    \State Update $\gamma = c* \gamma$;
\Until{the constraint violation falls below $\delta_1 > 0$}
\label{alg:capp1}
\end{algorithmic}
\end{algorithm}

\begin{algorithm}[t]
\caption{AO Algorithm for Solving Problem (25)}
\begin{algorithmic}[1]
\State Initial values for $\{\mathbf{Q}_k^{(0)}\}$ and set the iteration index $L{=}0$; set tolerance $\delta_0>0$ and the maximum number of iterations $L_{\max}$.
\Repeat
    \State  For given $\{\mathbf{Q}_k^{(L)}\}$, solve problem (26) to obtain the optimized $\{\mathbf{W}_k^{(L)}\}$ .
    \State  Update $\{\mathbf{Q}_k^{(L+1)}\}$ by solving problem (31) via \textbf{Algorithm 1}.
    \State  $L{=}L+1$.
\Until{the increase of the objective value of problem (25) is below $\delta_0$ or $L=L_{\max}$}
\label{alg:capp2}
\end{algorithmic}
\end{algorithm}

\subsection{Complexity Analysis}
The solution procedure for problem (25) is summarized in \textbf{Algorithm 1} and \textbf{Algorithm 2}. In each outer AO iteration, problems (26) and (31) are solved alternately: the former updates the active beamformers $\{\mathbf{W}_k\}$, while the latter updates the passive STAR-RIS variables $\{\mathbf{Q}_k\}$.
To enforce the rank-one structure of the STAR-RIS configuration, Algorithm~1 adopts a penalty mechanism in its inner loop; the penalty parameter is gradually increased until the constraint violation becomes negligible. Both subproblems are standard SDPs and are handled by an interior-point method.

Let $K_{\mathrm{IR}},\, K_{\mathrm{ER}},\, K_{\mathrm{Eve}}$ be the number of information users, energy-harvesting users, and eavesdroppers, and set $K_{\mathrm{sum}} = K_{\mathrm{IR}} + K_{\mathrm{ER}} + K_{\mathrm{Eve}}$. According to the analysis in \cite{5447068}, 
the approximate computational complexities for active and passive optimizations are $\mathcal{O}_{\mathrm{Active}} = \mathcal{O}\!\left(K_{sum}\left(M^{3.5}N_t^{3.5}+N_t^{3.5}\right)\right)$, and $\mathcal{O}_{\mathrm{Passive}} = \mathcal{O}\!\left(K_{sum}\left(M^{3.5}N_t^{3.5}+2M^{3.5}\right)\right)$.
Let $I_{\mathrm{in}}$ and $I_{\mathrm{out}}$ denote the inner and outer loop iteration counts. The overall complexity is $\mathcal{O}\!\left(I_{\mathrm{out}}\big(\mathcal{O}_{\mathrm{Active}}+I_{\mathrm{in}}\,\mathcal{O}_{\mathrm{Passive}}\big)\right)$. Since the objective value of problem (25) is non-decreasing across AO iterations and and remains upper bounded \cite{10304608}, the proposed method is ensured to converge to a stationary point.

\section{Simulation Results }
In this section, we assess the performance of the proposed STAR-RIS-aided LEO satellite SWIPT system and benchmark it against baseline schemes. For each simulation, the positions and channel states of all \text{Eves}, \text{IRs}, and \text{EHRs} were documented. The final performance metrics are an average taken from over 100 of these independent trials. In the considered simulation setup, the LEO satellite is $340$~km from the STAR-RIS, while all legitimate terminals, including both IRs and EHRs, are deployed at a distance of $10$~m from it. The eavesdroppers are randomly distributed within a $10$m radius around the STAR-RIS. Unless otherwise specified, the simulation parameters are set as follows: number of LEO satellite antennas $N_t = 4$, number of STAR-RIS elements $M = 20$, maximum transmit power $P_s = 20$~dBW, minimum rate requirement $R_{th}^{IR}= 1$~bits/s/Hz, $R_{th}^{Eve}= 1$~bits/s/Hz and noise powers $\sigma_k^2 = \sigma_e^2 = -170$~dBm. In this work, the proposed STAR-RIS-assisted LEO satellite network operates at 17.7 GHz (Ka band). The antenna gains are set to $G_{\mathrm{l}}=34~\text{dBi}$, $G_{\mathrm{l},s}=3.5~\text{dBi}$, and $G_{\max}=52~\text{dBi}$.

All wireless links between STAR-RIS and all users of IRs, EHRs, and Eves are modeled using Rician fading, capturing the presence of a dominant LoS path, which is expressed as \cite{9570143}:
\begin{equation}
\mathbf{h}_{k} = \sqrt{\chi_{h}(d)} \left( \sqrt{\frac{K_1}{K_1+1}} \mathbf{h}^{\text{LOS}}_{k} + \sqrt{\frac{1}{K_1+1}} \mathbf{h}^{\text{NLOS}}_{k}\right),    
\end{equation}
\begin{equation}
 \mathbf{g}_{k} = \sqrt{\chi_{g}(d)} \left(\sqrt{\frac{K_2}{K_2+1}} \mathbf{g}^{\text{LOS}}_{k} + \sqrt{\frac{1}{K_2+1}} \mathbf{g}^{\text{NLOS}}_{k}\right),   
\end{equation}

\begin{equation}
\mathbf{v}_{k} =\sqrt{\chi_{v}(d)} \left( \sqrt{\frac{K_3}{K_3+1}} \mathbf{v}^{\text{LOS}}_{k} + \sqrt{\frac{1}{K_3+1}} \mathbf{v}^{\text{NLOS}}\right),    
\end{equation}
where $\chi_{h}(d) = \dfrac{\chi_{0}}{d_{h}^{\alpha}}$, 
$\chi_{g}(d) = \dfrac{\chi_{0}}{d_{g}^{\alpha}}$, and 
$\chi_{v}(d) = \dfrac{\chi_{0}}{d_{v}^{\alpha}}$, 
in which $\chi_{0}$ denotes the path loss at the reference distance of $1$~m, 
$d_{h}$, $d_{g}$, and $d_{v}$ represent the distances from the STAR-RIS to the \text{IR}, \text{EHR}, and $k$-th \text{Eve}, respectively, and the corresponding path-loss exponent is $\alpha = 2.2$. The Rician factors are assumed identical for all links, with $K_{1} = K_{2} = K_{3} = 5$ dB. Here, $\mathbf{h}_k^{\mathrm{LOS}}$, $\mathbf{g}_k^{\mathrm{LOS}}$, and $\mathbf{v}_k^{\mathrm{LOS}}$ denote the deterministic LOS components, while $\mathbf{h}_k^{\mathrm{NLOS}}$, $\mathbf{g}_k^{\mathrm{NLOS}}$, and $\mathbf{v}_k^{\mathrm{NLOS}}$ represent the NLOS components. All LOS and NLOS components are modeled as i.i.d. complex Gaussian random variables.
In addition, the normalized maximum channel estimation errors of the $k$-th IR, ER, and Eve are set to be
$\rho_{H} = \frac{\epsilon_{1,k}}{\|\widehat{\mathbf{H}}_{k}\|}$, 
$\rho_{G} = \frac{\epsilon_{2,k}}{\|\widehat{\mathbf{G}}_{k}\|}$, 
and $\rho_{V} = \frac{\epsilon_{3,k}}{\|\widehat{\mathbf{V}}_{k}\|}$, respectively.
The uncertain degree of estimation error $\rho_{H} = \rho_{G} = \rho_{V}=\rho$. In the simulations, the tolerance parameters in Algorithms~1 and~2 are set to $\delta_{0} = 10^{-5}$, $\delta_{1} = 10^{-7}$, and $\delta_{2} = 10^{-7}$.

\begin{figure}[t!]
\centering
\includegraphics[width=0.45\textwidth]{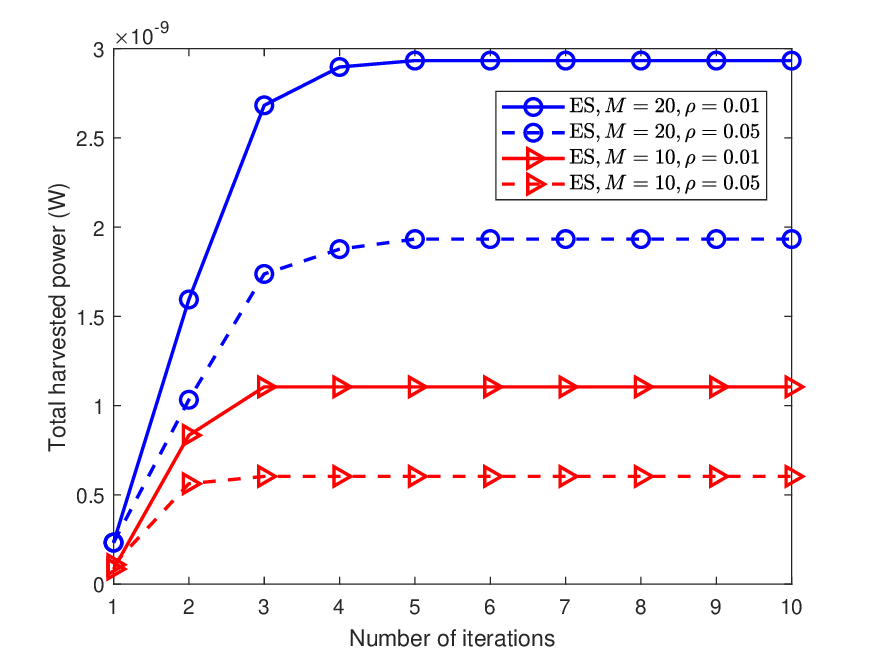}
    \caption{Convergence behavior of AO algorithm.}
    \label{fig:2}
\end{figure}

\begin{figure}[t!]
\centering
\includegraphics[width=0.45\textwidth]{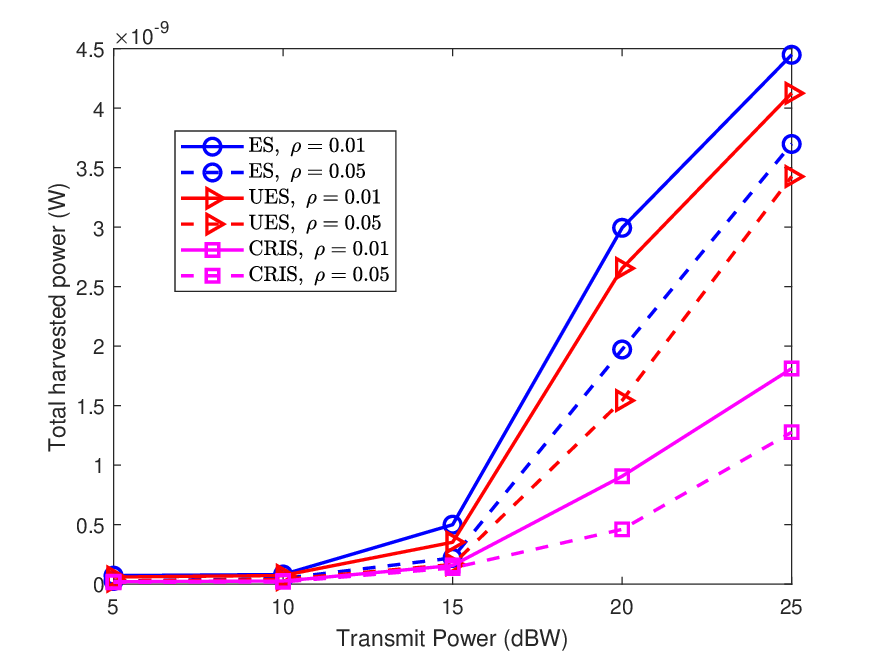}
    \caption{Total harvested power vs transmitted power.}
    \label{fig:3}
\end{figure}

To evaluate the performance gains offered by the proposed STAR-RIS-aided SWIPT system, two baseline schemes are considered for comparison. \textbf{Baseline 1: conventional RIS (CRIS)}, Where a transmit-only RIS and a reflect-only RIS are placed side by side at the STAR-RIS location, each with $M/2$ elements. \textbf{Baseline 2: uniform energy splitting  (UES)}, where in this scheme, all STAR-RIS elements employ identical amplitude coefficients for both transmission and reflection, with  $\beta_{m}^{T} = \beta^{T}, \quad \beta_{m}^{R} = \beta^{R}, \quad \forall m \in \mathcal{M}$,
where $\beta^{T}, \beta^{R} \in [0, 1]$ and $\beta^{T} + \beta^{R} = 1$. The resulting optimization problems for both baseline schemes can be solved using the same procedures as those described for the proposed method by applying \textbf{Algorithm 1} and  \textbf{2}.

In Fig. \ref{fig:2}, we illustrate the convergence behavior of the proposed AO algorithm for the STAR-RIS operating under the ES protocol, considering different number of elements $M=(10,20)$ and uncertainty levels $\rho=(0.01,0.05)$. For a representative channel realization, the total harvested power increases rapidly with the iteration index and then stabilizes. A larger number of STAR-RIS elements $M=20$ achieves a higher steady-state harvested power, and also results in a slightly higher number of iterations required for convergence, while a larger uncertainty level $\rho=0.05$ yields a lower limiting value due to more conservative robust constraints. These results confirm the monotone and fast convergence of the proposed AO procedure.

In Fig.\ref{fig:3}, we present the total harvested power versus the transmit power. As expected, the total harvested power increases with the transmit power for all schemes. The ES protocol consistently outperforms the UES and CRIS schemes due to its ability to independently adjust the amplitude and phase shift for each element, allowing more flexible resource allocation compared to other schemes. Moreover, For both the proposed and baseline designs, a smaller uncertainty level (\(\rho=0.01\)) yields higher harvested power than \(\rho=0.05\), reflecting the reduced conservatism of the robust constraints. The curves are close at low power (up to about 15 dBW), but the harvested power and the gap between ES and the baselines grow markedly at higher power (20-25 dBW), which is consistent with the large path loss over long links.

\begin{figure}[t!]
\centering
\includegraphics[width=0.45\textwidth]{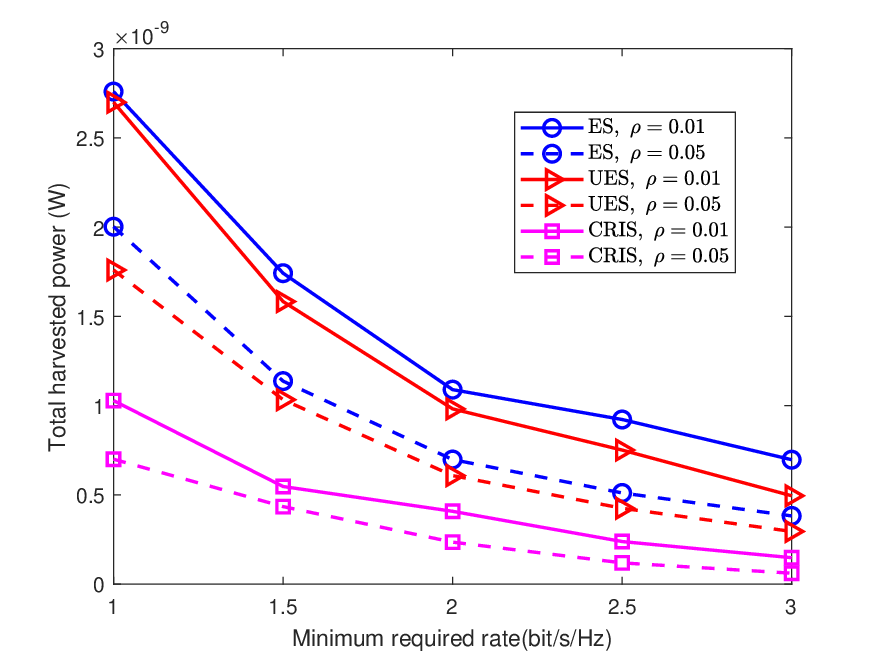}
    \caption{Total harvested power vs required rate at IR.}
    \label{fig:4}
\end{figure}

In Fig. \ref{fig:4}, the total harvested power is shown as a function of the minimum required rate at the IR. Meeting stricter rate targets diverts more transmit power to information delivery, leaving less for energy transfer. The proposed ES design attains the highest harvested power across the entire range, owing to its per-element amplitude–phase control, whereas UES and CRIS are limited by fewer degrees of freedom. A lower channel uncertainty level enhances harvested power by enabling less conservative beamforming, whereas higher uncertainty necessitates more cautious power allocation. The performance disparity across the schemes is accentuated in the presence of low rate requirements and gradually narrows as the rate constraint becomes more dominant at higher rate demands.

\begin{figure}[t!]
\centering
\includegraphics[width=0.45\textwidth]{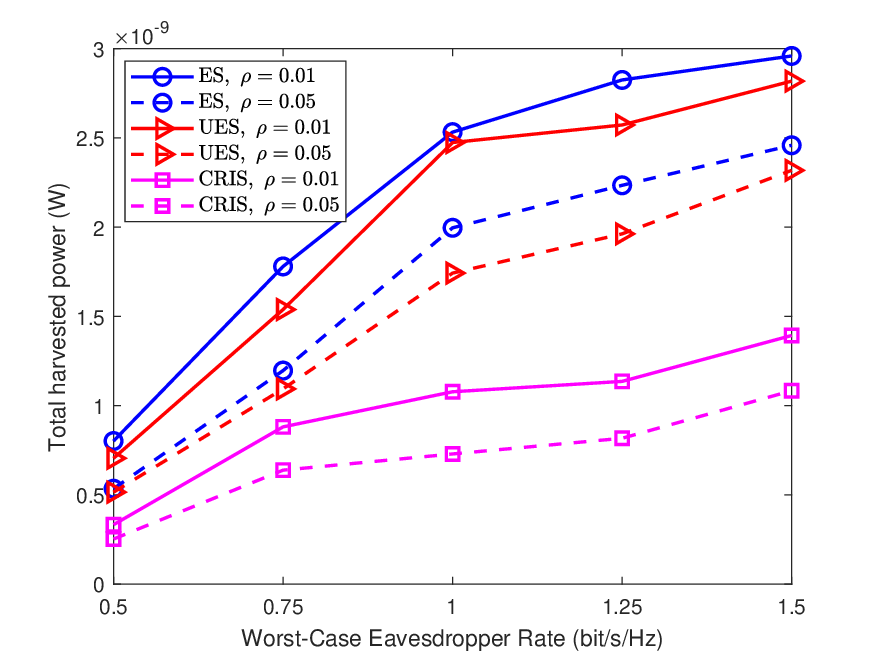}
    \caption{Total harvested power vs required rate at Eave.}
    \label{fig:5}
\end{figure}

In Fig. \ref{fig:5}, the total harvested power is plotted against the worst-case eavesdropper rate. Raising the eavesdropper-rate threshold relaxes the secrecy constraint, enabling more transmit power to be allocated to energy harvesting and thereby raising the harvested power across all schemes. The proposed model consistently achieves the highest harvested power across the entire range. Moreover, Larger channel estimation errors force more power to meet  QoS; with a fixed budget, beamforming becomes more conservative and harvested power drops.

\begin{figure}[t!]
\centering
\includegraphics[width=0.45\textwidth]{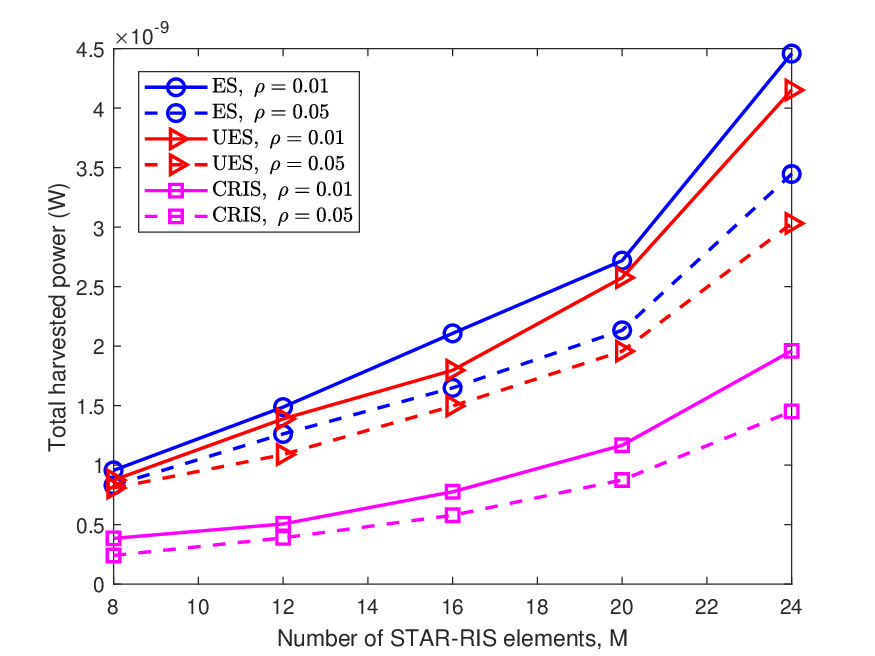}
    \caption{Total harvested power vs the number of elements at STAR-RIS.}
    \label{fig:6}
\end{figure}

In Fig. \ref{fig:6}, we plot the total harvested power versus the number of STAR-RIS elements M. Harvested power increases with M for all schemes because a larger aperture and more degrees of freedom enable tighter phase alignment and stronger energy focusing, which mitigates long-distance attenuation. The proposed model attains the highest harvested power compare to UES and CRIS. Channel uncertainty also matters where the curve with $\rho=0.01$ remains above the one with $\rho=0.05$, and the gap tends to widen as M grows since estimation errors in the cascaded links accumulate with more elements, forcing more conservative beamforming and reserving more power to maintain rate and secrecy quality of service, which limits the harvestable energy.

\begin{figure}[t!]
\centering
\includegraphics[width=0.45\textwidth]{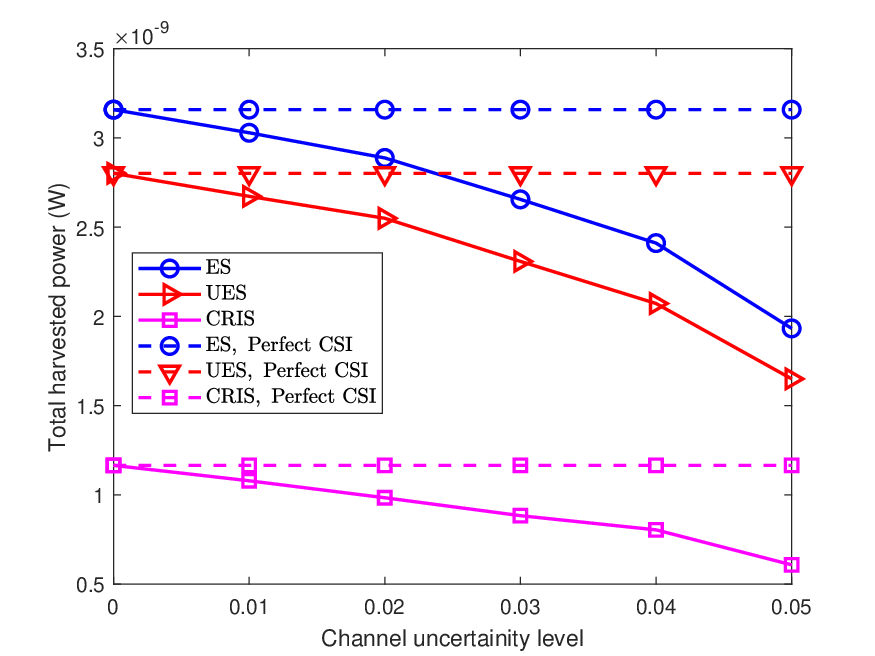}
    \caption{Total harvested power vs channel uncertainty level.}
    \label{fig:7}
\end{figure}

In Fig. \ref{fig:7}, we plot the total harvested power against the channel-uncertainty level $\rho$. All curves drop as $\rho$ increases because channel errors affect the IR links more, so more transmit power is needed to satisfy the IR rate requirement and less remains for harvesting. The proposed ES scheme achieves the highest harvested power over the range. Its decline with larger $\rho$ is steeper than that of the conventional RIS because the many controllable elements of the STAR-RIS amplify cascaded-channel estimation errors. UES lies between ES and CRIS. The dashed Perfect-CSI curves are optimistic upper bounds since they ignore estimation errors.

\begin{figure}[t!]
\centering
\includegraphics[width=0.45\textwidth]{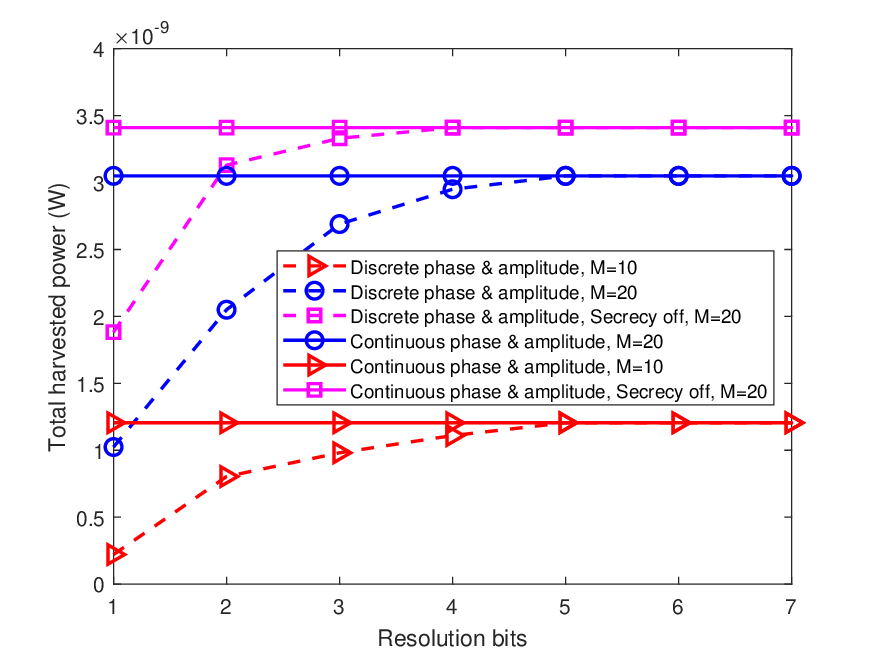}
    \caption{Total harvested power versus the number
 of phase shift and amplitude quantization bits.}
    \label{fig:8}
\end{figure}

To assess practical implementation trade-offs, we examine in Fig. \ref{fig:8} how the number of quantization bits at the STAR-RIS affects the harvested power. In the discrete case, the phase values are drawn from
$\left\{0, \frac{2\pi}{2^q}, \ldots, \frac{2\pi(2^q-1)}{2^q}\right\}$,
and the amplitude values from
$\left\{0, \frac{1}{2^q-1}, \frac{2}{2^q-1}, \ldots, 1\right\}$,
selected to be closest to the corresponding continuous values. We compare discrete designs with joint phase and amplitude quantization against their continuous-control upper bounds, under both secrecy-enabled and secrecy-off settings, and for $M=10$ and $M=20$.  Harvested power increases with q as finer per-element control improves energy focusing, but the gains saturate beyond \(q=5\). With secrecy enforced, \(q=4\) already achieves more than \(98\%\) of the continuous benchmark, and additional bits yield only marginal improvement at higher hardware cost. When the secrecy (worst-case eavesdropper-rate) constraint is removed, near-continuous performance is reached with fewer bits. Importantly, with 5-bit quantization, the performance gap between the discrete design and the continuous benchmark is negligible, independent of the STAR-RIS element count, indicating that moderate-resolution implementations offer a favorable balance
between performance and complexity.

\section{Conclusion}
In this work, we studied a STAR-RIS-aided SWIPT system in a LEO satellite network with blocked direct links. We formulated a joint beamforming and STAR-RIS coefficient optimization problem to maximize the minimum harvested power under secrecy and power constraints. To solve the resulting non-convex problem, we employed an alternating optimization framework combined with a penalty-based method, enabling efficient handling of rank-one constraints.
Simulation results revealed that increasing the secrecy rate threshold for users and enlarging the number of STAR-RIS elements can degrade the harvested energy performance due to higher channel estimation errors. In contrast, increasing the transmit power significantly enhances harvested energy, highlighting the trade-off between system security, hardware configuration, and energy transfer efficiency. Overall, the proposed approach outperforms conventional RIS and baseline schemes, demonstrating the potential of STAR-RIS for secure and sustainable satellite-assisted wireless networks.

\section*{APPENDIX A\\Proof of Theorem 1}

The relaxed active-beamforming subproblem in \eqref{eq:25} is convex in the variables $\{\mathbf{W_k}\}$ and satisfies Slater's condition; hence, strong duality holds. We therefore consider the Lagrangian of \eqref{eq:25}. Let $\mathbf{U_{1}}$, $\mathbf{U_{2,k}}$, and $\mathbf{U_{3,k}}$ denote the LMI blocks generated by the S-procedure in \eqref{eq:22}-\eqref{eq:24}, respectively. The associated dual variables are $\lambda\ge 0$, $\mathbf{S_1}\succeq 0$ for the IR rate constraint\eqref{eq:22}, $\mathbf{S_{2,k}}\succeq 0$ for the energy harvest constraint  \eqref{eq:23}, and $\mathbf{S_{3,k}}\succeq 0$ for the Eve leakage rate constraint\eqref{eq:24}, and $\mathbf{Y_k}\succeq 0$ for the cone constraint $\mathbf{W_k}\succeq 0$. The remaining terms that play no role in the proof are aggregated into $\boldsymbol{\Upsilon}$.

\begin{align}
L
&= \xi - \lambda\!\left(\sum_{k}\operatorname{Tr}(\mathbf{W}_k)-P_{\max}\right)
   + \sum_{k}\operatorname{Tr}(\mathbf{Y}_k\,\mathbf{W}_k) \notag\\
&\quad + \sum_{k\in\{t,r\}}\operatorname{Tr}\!\big(\mathbf{U}_{1}\,\mathbf{S}_{1}\big)
   + \sum_{k\in\{t,r\}}\operatorname{Tr}\!\big(\mathbf{U}_{2,k}\,\mathbf{S}_{2,k}\big) \notag\\
&\quad + \sum_{k\in\{t,r\}}\operatorname{Tr}\!\big(\mathbf{U}_{3,k}\,\mathbf{S}_{3,k}\big)
   + \boldsymbol{\Upsilon}. \label{eq:36}
\end{align}

We now turn to the Karush--Kuhn--Tucker (KKT) conditions with respect to $W_k$.
\begin{subequations}\label{eq:37}
\begin{align}
&\lambda^\star \ge 0,\quad \mathbf{S}_{1}^\star,\ \mathbf{S}_{2,k}^\star,\ \mathbf{S}_{3,k}^\star,\ \mathbf{Y}_k^\star \succeq 0, \label{eq:37a}\\
&\mathbf{Y}_k^\star\,\mathbf{W}_k^\star = 0,\qquad \forall\,k\in\{t,r\}. \label{eq:37b}
\end{align}

\begin{align}
\mathbf{Y}_k^\star
&= \lambda^\star \mathbf{I}_{N_t} \notag\\
&\quad - \left(\sum_{k\in\{t,r\}}\sum_{m=1}^{M}
   \Big[\mathrm{vec}(\hat{\mathbf{H}}_k)^{\!H}\,\big(\mathbf{U}_{1}^\star\big)\,\frac{1}{\Gamma}\,\mathrm{vec}(\hat{\mathbf{H}}_k)\Big]_{x:y,\,\hat{x}:\hat{y}}\right) \notag\\
&\quad + \sum_{k\in\{t,r\}}\sum_{m=1}^{M}
   \Big[\mathrm{vec}(\hat{\mathbf{H}}_k)^{\!H}\,\big(\mathbf{U}_{1}^\star\big)\,\mathrm{vec}(\hat{\mathbf{H}}_k)\Big]_{x:y,\,\hat{x}:\hat{y}} \notag\\
&\quad - \sum_{k}\sum_{m=1}^{M}
   \Big[\mathrm{vec}(\hat{\mathbf{G}}_k)^{\!H}\,\big(\mathbf{U}_{2,k}^\star\big)\,\mathrm{vec}(\hat{\mathbf{G}}_k)\Big]_{x:y,\,\hat{x}:\hat{y}} \notag\\
&\quad - \sum_{k\in\{t,r\}}\sum_{m=1}^{M}
   \Big[\mathrm{vec}(\hat{\mathbf{V}}_k)^{\!H}\,\big(\mathbf{U}_{3,k}^\star\big)\,\mathrm{vec}(\hat{\mathbf{V}}_k)\Big]_{x:y,\,\hat{x}:\hat{y}} \notag\\
&\quad + \sum_{k\in\{t,r\}}\sum_{m=1}^{M}
   \Big[\mathrm{vec}(\hat{\mathbf{V}}_k)^{\!H}\,\big(\mathbf{U}_{3,k}^\star\big)\,\frac{1}{\mathbf{\eta}}\,\mathrm{vec}(\hat{\mathbf{V}}_k)\Big]_{x:y,\,\hat{x}:\hat{y}},
\label{eq:37c}
\end{align}
\end{subequations}

where \(x=\hat{x}=(m-1)N_t+1\) and \(y=\hat{y}=mN_t\), where \(m=1,\ldots,M\) indexes the STAR-RIS elements and \(N_t\) is the number of satellite transmit antennas. From complementary slackness \eqref{eq:37b}, the column space of the optimal $\mathbf{W}_k^\star$ lies entirely within the null space of the optimal $\mathbf{Y}_k^\star$, which links \(\operatorname{rank}(W_k^{\star})\) to the rank of $\mathbf{Y}_k^\star$. Under any feasible nontrivial setting with \(P_{\max}>0\), \(\Gamma>0\), and \(\eta>0\), we necessarily have \(\lambda^{\star}>0\) and \(W_k^{\star}\neq 0\). Invoking the standard rank-reduction result for this SDP structure then yields \(\operatorname{rank}(W_k^{\star})=1\) for all \(k\). A detailed proof for this property in similar problems can be found in \cite{5447068}.

%\textbf{Theorem 1:} The optimal solution of the relaxed version of problem~(19a), where the rank-one constraints are omitted, always satisfies $\operatorname{Rank}(\mathbf{W}_i^*) = 1, \forall i \in \mathcal{K}_I$ and $\operatorname{Rank}(\mathbf{V}_j^*) \leq 1, \forall j \in \mathcal{K}_E$, under the conditions $P_{\max} > 0$ and $\gamma_i > 0$.

%%====> References <===%%
\bibliographystyle{IEEEtran}
\bibliography{Secure_STAR_RIS}

\end{document}